\documentclass[12pt]{article}
\usepackage{amssymb}
\usepackage{amsmath}

\newtheorem{theorem}{Theorem}

\newtheorem{corollary}[theorem]{Corollary}

\begin{document}
\title
{Inverse boundary value problem for Schr\"odinger equation in
cylindrical domain by partial boundary data}

\author{
O.~Yu.~Imanuvilov\thanks{ Department of Mathematics, Colorado State
University, 101 Weber Building, Fort Collins, CO 80523-1874, U.S.A.
E-mail: {\tt oleg@math.colostate.edu}. }\, and \,
M.~Yamamoto\thanks{ Department of Mathematical Sciences, University
of Tokyo, Komaba, Meguro, Tokyo 153, Japan e-mail:
myama@ms.u-tokyo.ac.jp} }

\date{}

\maketitle
\begin{abstract}
Let $\Omega\subset \Bbb R^2$ be a bounded domain with
$\partial\Omega\in C^\infty$ and $L$ be a positive number. For a
three dimensional cylindrical domain $Q=\Omega\times (0,L)$, we
obtain some uniqueness result of determining a complex-valued
potential for the Schr\"odinger equation from partial Cauchy data
when Dirichlet data vanish on a subboundary
$(\partial\Omega\setminus\widetilde{\Gamma}) \times [0,L]$ and the
corresponding Neumann data are observed on $\widetilde\Gamma \times
[0,L]$, where $\widetilde\Gamma$ is an arbitrary fixed open set of
$\partial\Omega.$
\end{abstract}

This article is concerned with the inverse boundary value problem of
determination of a complex-valued potential for the Schr\"odinger
equation in a cylindrical domain from partial boundary data. More
precisely the problem is as follow. Let $Q=\Omega\times (0,L)$,
where $\Omega\subset \Bbb R^2$ is a bounded domain with
$\partial\Omega\in C^\infty.$ Let $\widetilde \Gamma$ be an
arbitrary, fixed subdomain of $\partial\Omega.$ Denote
$\Gamma_0=Int(\partial\Omega\setminus\widetilde \Gamma)$,
$\widetilde \Sigma=\widetilde \Gamma\times [0,L]$ and
$\Sigma_0=\Gamma_0\times [0,L]$.

In $Q$, we consider the Schr\"odinger equation with some
complex-valued potential $q$:
\begin{equation}\label{popo1}
L_q(x,D)u=(\Delta +q)u=0\quad\mbox{in}\,\, Q.
\end{equation}

Consider the following Dirichlet-to-Neumann map
$\Lambda_{q,\Sigma_0}$
\begin{equation}\label{popo}
\Lambda_{q,\Sigma_0} f= \frac{\partial u} {\partial
\nu}\vert_{\partial Q\setminus\Sigma_0},\quad \mbox{where}\,\,
L_q(x,D)u=0\quad\mbox{in}\,\,Q,\quad u\vert_{\Sigma_0}=0,\quad
u\vert_{\partial Q\setminus\Sigma_0} =f
\end{equation}
with domain $$ D(\Lambda_{q,\Sigma_0})=\{f\in H^\frac 12(\partial
Q)\vert \mbox{supp} f\subset \partial Q\setminus \Sigma_0,\,\,\,
(f,g)_{L^2(\partial Q)}=0 \quad\forall g\in \mathcal N\}
$$
and
$$
\mathcal N=\{g\vert L_q(x,D)u=0\quad\mbox{in}\,\,Q, \quad
u\vert_{\partial Q}=0, \quad \frac{\partial
u}{\partial\nu}\vert_{\partial Q} =g\}.
$$
 The problem (\ref{popo1}) and (\ref{popo}) is the
generalization of the inverse boundary value problem of recovery of
the conductivity, which is also known as Calder\'on's problem (see
\cite{C}). It is related to many practical applications, for
example, detecting oil or minerals by applying voltage and measuring
the fluxes on earth's surface. See also Cheney, Issacson and Newell
\cite{CIN} for applications to medical imaging of EIT.

In case when $Q$ is a general domain in $\Bbb R^n$ with $n \ge 2$,
$\Sigma_0=\emptyset$  (i.e., the case of full Dirichlet-to-Neumann
map), the unique recovery of the conductivity  was established in
\cite{N} and \cite{SU}  in two and three dimensional cases
respectively. From the practical point of view, the assumption
$\Sigma_0=\emptyset$ ,  means that one has to set up voltages and
measure the fluxes on the whole boundary is very restrictive. In
practice, this assumption does not often hold, for example because
the domain $Q$ is extremely large or we can not access to some part
of $\partial Q$, e.g.,  the domain has cavities located inside. For
the inverse boundary value problem with such partial
Dirichlet-to-Neumann map, we refer to the following works. In
\cite{BuU} Bukhgeim and Uhlmann show that if voltages are applied on
the boundary $\partial Q_-$ and the corresponding fluxes are
measured on some part $\partial Q_+$ which is approximately equal to
$\partial Q\setminus \partial Q_-$, then the potential can be
uniquely determined. This result and a recent improved result
\cite{KSU} still require  the access to the whole boundary $\partial
Q.$ In \cite{I},  Isakov  solves the case where voltage applied and
current measured  on the same set $\partial Q_-$ provided that
subboundary $\partial Q\setminus \partial Q_-$ is a part of some
sphere or some plane.  All the above mentioned papers treat the case
where the spatial dimension more or equal $ 3$.

The purpose of this article is to establish the uniqueness with weak
constraints on such subboundary in the case of three dimensional
cylindrical domain $Q$. Our proof is based on the Radon transform
and gives the uniqueness in higher dimensional domains with
generalized geometrical configurations (not necessarily cylindrical
domains).

As for related work in slabs, see Ikehata \cite{Ik}, Krupchyk,
Lassas and Uhlmann \cite{KLU}, Uhlmann  {\it et al} \cite{LG}.   See
Novikov \cite {Nov} for conditional stability results for Calderon's
problem.

For general two dimensional domain, \cite{IUY} proved the unique
recovery of a potential for the Schr\"odinger equation in the case
when voltage applied  and flux measured  both on an arbitrary open
set of $\partial Q$. Thus \cite{IUY} established the best possible
uniqueness in the two dimensional case with data
$\Lambda_{q,\Sigma_0} $ defined by (\ref{popo}). Also see
\cite{IUY1} which deals with the same inverse problem for more
general second-order elliptic equations in the two dimensional case
and \cite{IY} improves the result of \cite{IUY} in terms of
regularity assumption of potential for the Schr\"odinger equation.

We introduce the subset $\mathcal O$ of domain $\Omega$
$$
\mathcal O=\Omega\setminus Ch(\bar\Gamma_0),\quad  Ch(\bar\Gamma_0)=
\{ x\vert x=\lambda x^1+(1-\lambda)x^2, x^1, x^2\in
\overline{\Gamma_0}, \lambda\in (0,1)\}.
$$

We have
\begin{theorem}\label{oppp}
Let  $q_1,q_2$ be Lipschitz functions. If
$\Lambda_{q_1,\Sigma_0}=\Lambda_{q_2,\Sigma_0}$ and
$D(\Lambda_{q_1,\Sigma_0})\subset D(\Lambda_{q_2,\Sigma_0})$, then
$q_1=q_2$ in $\mathcal O\times [0,L]$.
\end{theorem}

From theorem \ref{oppp} we obtain immediately
\begin{corollary} Let $\Omega$ be concave near $\Gamma_0$ and
potentials   $q_1,q_2$ be Lipschitz functions such that
$\Lambda_{q_1,\Sigma_0}=\Lambda_{q_2,\Sigma_0}$ and
$D(\Lambda_{q_1,\Sigma_0})\subset D(\Lambda_{q_2,\Sigma_0}).$ Then
$q_1=q_2$ in $Q$.
 \end{corollary}

 First we formulate the following Carleman estimate  with the linear weight function
 $\varphi=x_3$ for the Schr\"odinger  operator (\ref{popo1}).
 Denote $\Vert \cdot\Vert_{H^{1,\tau}(Q)}=\Vert \cdot\Vert_{H^1(Q)}+\vert\tau \vert\Vert \cdot\Vert_{L^2(Q)}.$ In
 \cite{BuU} the following theorem is proved:

 \begin{theorem}\label{zombi} Let $q\in L^\infty(Q).$ There exist $\tau_0$ and constant $C$ independent of $\tau$ such that for all $\tau\ge\tau_0$
 \begin{equation}\label{solnishko}
 \Vert u e^{\tau\varphi}\Vert_{H^{1,\tau}(Q)}\le C (\Vert (L_q(x,D) u) e^{\tau\varphi}\Vert_{L^2(Q)}+\root\of{\tau}\Vert (\frac{\partial u}{\partial\nu}
 e^{\tau\varphi})(\cdot,L)\Vert_{L^2(\Omega)}) \quad \forall u\in
 H^1_0(Q).
 \end{equation}
 \end{theorem}

 Next we formulate some known   results on the generalized Radon transform:
 $$
 (\mathcal R_\mu f)(\omega, p)=\int_{<\omega, x>=p}f(x) e^{\mu<\omega^\bot,
 x>}ds, \quad (\omega,p)\in S^1\times \Bbb R , \quad \omega^\bot=(\omega_2,-\omega_1).
 $$
 The following theorem  is proved in \cite{Kur}
 \begin{theorem} Let $f$ be a Lipschitz function with compact
 support. If $(\mathcal R_\mu f)(\omega, p)=0$ for all $p>r$ then
 $\mbox{supp}\, f\subset \{x\in \Bbb R^2\vert \vert x\vert\le r\}.$
 \end{theorem}

Similar to Corollary 2.8 of \cite{Helg} p. 14 we prove
\begin{corollary}\label{zebra1} Let $f$ be a Lipschitz function in $\Bbb R^2$ with
compact support and $E$ be a bounded convex set in $\Bbb R^2.$ If
$(\mathcal R_\mu f)(\omega, p)=0$ for all lines $<\omega,\tilde
x>=p$ which do not intersect $E$ then
$$
f(x)=0\quad \forall x\notin E.
$$
\end{corollary}
 {\bf Proof of
Theorem \ref{oppp}.} Let point $\hat x=(\hat x_1, \hat x_2, \hat
x_3)\in \mathcal O\times (0,L).$ Since $Ch(\bar\Gamma_0)$ is the
convex closed set the point $(\hat x_1,\hat x_2)$ can be separated
from it by some line $\ell.$ Then the line which is parallel to
$\ell$ and
 pass through  $(\hat x_1,\hat x_2)$ does not intersect  $Ch(\bar\Gamma_0).$ After possible rotation and translation we may assume that $\hat
x_1=0, \hat x_2>0$  and axis $x_2$ does not intersect
$Ch(\bar\Gamma_0).$ Therefore it suffices to prove that
$$ q_1=q_2\quad \mbox{on } Q\cap \{x\vert x_3\in [0,L], x_1=0\}. $$
Without loss of generality we may assume that $\Omega\subset \Bbb
R^1_+.$ Let $m$ be a smooth function defined on $\Bbb R^1$  such
that $\vert m'\vert <1.$ Denote $\nabla
'=(\partial_{x_1},\partial_{x_2}).$ Consider the eiconal equation
 \begin{equation}
\label{vovochka} \vert\nabla '
\Psi\vert=1\quad\mbox{in}\,\,\Omega,\quad \Psi\vert_{x_2=0}=m.
\end{equation}
This equation can be integrated by the method of characteristics.
The solutions, as long as they exist,  have the  following form
\begin{equation}\label{zavtra}
\Psi(x_0+t\alpha(x_0)\vec e_1 +t\beta(x_0)\vec e_2)=m(x_0)+ t \quad
\forall x_0\in \{x'\vert x_2=0\},\quad t>0,
\end{equation}
where $\alpha(x_0)= m'(x_0),\beta(x_0)=\root\of{1-\alpha^2(x_0)}.$

Next construct the function $\Psi$  more explicitly using the
implicit function theorem. Consider the following mapping:
$$
F(y)=y_2\alpha(y_1)\vec e_1+y_2\beta(y_1)\vec e_2+\left
(\begin{matrix} y_1\\0\end{matrix}\right ),\quad y=(y_1,y_2).
$$

Assume that
\begin{equation}\label{zombi}
\alpha(0)= m'(0)=0.
\end{equation}
Then \begin{equation}\label{inn} F(0,t)=(0,t) \end{equation} and

$$F'(y)=(y_2(\alpha'(y_1)\vec e_1 +\beta'(y_1)\vec e_2)+\left (\begin{matrix}  1\\0 \end{matrix}\right
),\alpha(y_1)\vec e_1
+\beta(y_1)\vec e_2)=\left(\begin{matrix}  y_2\alpha'+1 &\alpha \\
 y_2\beta' &\beta
\end{matrix}\right ).
$$
In particular

$$
F'(0,y_2)=\left (\begin{matrix}  1 +y_2\alpha'(0) & 0\\ 0 &
1\end{matrix} \right ).$$ As long as  the function $1+y_2\alpha'(0)$
is positive, there exists the inverse matrix

\begin{equation}\label{zinka1} (F')^{-1}(0,y_2)=\left (\begin{matrix} \frac{1}{1
+y_2\alpha'(0)}& 0\\ 0 &1
\end{matrix} \right ).
\end{equation}
Let $K$ be  a positive number such that
$$
\Omega\cap \{ x'\vert x_1=0\}\subset \{ x'\vert x_1=0, 0< x_2 < K\}.
$$
By (\ref{zinka1}) there exists $\epsilon (K)$  for any $\alpha\in
{\cal X}=\{ \phi\in C^2_0(-1,1)\vert \phi(0)=0\}$ such that $\Vert
\alpha\Vert_{C^2[-1,1]}\le \epsilon(K)$ there exists $\tilde\delta
>0$ such that on the set $[-\tilde\delta,\tilde\delta]\times [0,2K]$ the matrix $(F')^{-1}$ is correctly defined
$$
(F')^{-1}(y)=\frac{1}{det F'(y)}\times\left(\begin{matrix}\beta  &-\alpha \\
 -y_2\beta' & y_2\alpha'+1
\end{matrix}\right ).
$$
Then by (\ref{inn}) and  the implicit function theorem  there exist
$\delta>0$ such that the mapping $x'\rightarrow y(x') $ is correctly
defined on $\frak D=[-\delta,\delta]\times [0,K]$ and the derivative
of this mapping given by formula:

\begin{equation}\label{L1}
\frac{\partial y}{\partial x'}=(F')^{-1}(y(x')).
\end{equation}
Differentiating the first columns on both sides of the matrix
equation (\ref{L1}) with respect to $x_1$ and using (\ref{zombi}) we
have
\begin{equation}\label{zinka}\left (\begin{matrix} \frac{\partial^2 y_1}{\partial x_1^2}\\
\frac{\partial^2 y_2}{\partial x_1^2}\end{matrix}\right )(0,x_2)
=-\frac{y_2\alpha''(0)}{(1+y_2\alpha'(0))^3}\left (\begin{matrix}
1\\0 \end{matrix}\right ) + \frac{1}{(1+y_2\alpha'(0))^2}\left
(\begin{matrix} 0\\-y_2(\alpha'(0))^2 \end{matrix}\right ).
\end{equation}

Then the function $\Psi$ can be determined by formula
$$
\Psi(x')=m_0(y_1(x'))+y_2(x').
$$
The  short computations and (\ref{zinka1}), (\ref{zinka}) imply
\begin{eqnarray}\label{9}
\Delta
\Psi(0,x_2)=m_0''(0)(\partial_{x_1}y_1)^2(0,x_2)+m_0'(0)(\partial^2_{x_1}y_1)(0,x_2)+\partial^2_{x_1}y_2(0,x_2)=
\frac{m_0''(0)}{(1 +y_2\alpha'(0))^2}\nonumber\\
+\frac{y_2(\alpha'(0))^2}{(1+y_2\alpha'(0))^2}= \frac{\alpha'(0)}{(1
+y_2\alpha'(0))^2}+\frac{y_2(\alpha'(0))^2}{(1+y_2\alpha'(0))^2}=
\frac{\alpha'(0)}{(1 +y_2\alpha'(0))}.
\end{eqnarray}

Let $a_0(x')$ be a function such that
\begin{equation}\label{10}
2(\nabla '\Psi,\nabla 'a_0)+\Delta_{x'} \Psi a_0=0\quad \mbox{in
}\,\,\frak O,
\end{equation}
 and $a(x')$ be a smooth function such that
 \begin{equation}\label{lora}
(\nabla '\Psi,\nabla 'a)=0\quad \mbox{in }\,\,\frak O.
\end{equation}

Next we construct the functions $a_0$ and $a$.

In order to construct the function $a(x')$   we take a smooth
function $r$
\begin{equation} \label{big}r\in C^\infty_0(-\epsilon,\epsilon),
\end{equation}
where $\epsilon$ is a small parameter and set:
\begin{equation}\label{big1}
a(x')=r(x_0) \quad \mbox{on the line}\,\,\,\{x'\in \Bbb R^2\vert
x'=\left (\begin{matrix}x_0\\0 \end{matrix}\right )+t\alpha(x_0)\vec
e_1 +t\beta(x_0)\vec e_2, t>0\}.
\end{equation} We  claim that for the function $a$
defined by these formula we have (\ref{lora}). Set $\vec v_1=
\alpha(x_0)\vec e_1 +\beta(x_0)\vec e_2, \vec v_2=\alpha(x_0)\vec
e_1 -\beta(x_0)\vec e_2.$ Then by (\ref{zavtra}) $\vert
\partial_{\vec v_1}\Psi\vert=1.$ Since $\vert \nabla '\Psi\vert=1$ we have that the vector $\vec
v_1$ is parallel to $\nabla '\Psi.$ Hence vectors $\vec v_j$ are
orthogonal. So
$$
\partial_{\vec v_2}\Psi=0.
$$
Therefore $$\partial_{\nabla '\Psi} a=\vert\nabla '\Psi\vert
\partial_{\frac{\nabla '\Psi}{\vert\nabla '\Psi\vert}} a
=\vert\nabla '\Psi\vert \partial_{\vec v_1} a=0.
$$
Hence the formula (\ref{lora}) is proved.

We integrate equation (\ref{10}) by the characteristic method. In
particular  using (\ref{9}) we have
\begin{equation}\label{zemlanika}
a_0(0,x_2)=e^{-\frac 12\int_0^{x_2}\frac{\alpha'(0)}{(1
+y_2\alpha'(0))}dy_2}=e^{-\frac 12 ln (1
+x_2\alpha'(0))}=\frac{1}{\root\of{1 +x_2\alpha'(0)}}.
\end{equation}

Next we construct the complex geometric optics solution
$u_1(x,\tau)$ for the Schr\"odinger operator with the potential
$q_1.$ For the principal term of complex geometric optics solution
we set
\begin{equation}
U=e^{(\tau+N)(x_3+i\Psi(x'))}aa_0.
\end{equation} The set $\mathcal O$ is closed and the axis $x_2$ does not intersect this set.
So there exists a neighborhood of the set $\{x'\vert x_2\in [0,K]\}$
such that it does not intersect $\mathcal O.$ Thanks to (\ref{big})
and (\ref{big1}), choosing a positive parameter $\epsilon$
sufficiently small, we obtain
\begin{equation}\label{lider}
U\vert_{\Sigma_0}=0.
\end{equation}
The simple computations imply
\begin{eqnarray}
L_{q_1}(x,D)U=(\tau+N)^2(\nabla (x_3+i\Psi),\nabla
(x_3+i\Psi))U\nonumber\\
+(\tau+N)(2(\nabla'\Psi,\nabla ' a_0)+\Delta_{x'} \Psi a_0) a
e^{(\tau+N)(x_3+i\Psi(x'))}\nonumber\\
+e^{(\tau+N)(x_3+i\Psi(x'))}\Delta_{x'} (a_0 a)+q_1U=
e^{(\tau+N)(x_3+i\Psi(x'))}\Delta_{x'}(a_0 a)+q_1U.
\end{eqnarray}

Observe that the  functions
$(e^{(\tau+N)(x_3+i\Psi(x'))}\Delta_{x'}(a_0 a)+q_1U)e^{-\tau x_3}$
are uniformly bounded in $\tau$ in norm of the space $ L^2(Q).$
Consequently using the results of \cite{BuU}
 we construct the last term in complex geometric optics solution- the function  $u_{cor}(\cdot,\tau)$ such that
 \begin{equation}\label{lider1} L_{q_1}(x,D)(e^{\tau\varphi}u_{cor})=-L_{q_1}(x,D)U\quad \mbox{in }\,\,Q,\quad
 u_{cor}\vert_{\Sigma_1}=0
 \end{equation}
 and
 \begin{equation}\label{cor}
 \Vert u_{cor}\Vert_{L^2(Q)}=O(\frac 1\tau)\quad \mbox{as}\,\,\tau\rightarrow +\infty.
 \end{equation}
 Hence, by (\ref{lider}), (\ref{lider1}) and (\ref{cor}) we have the representation
\begin{equation}\label{lora1}
u_1=U+e^{\tau x_3}O_{L^2(Q)}(\frac 1\tau)\quad\mbox{as}\quad
\tau\rightarrow +\infty, \,\,u_1\vert_{\Sigma_0}=0.
\end{equation}
(By $O_{L^2(Q)}$ we mean any function $f(\cdot,\tau)$ such that
 $\Vert f(\cdot,\tau)\Vert_{L^2(Q)}=O(1)$ as $\tau\rightarrow +\infty.$ ) Similarly we set
\begin{equation}\label{ol}
V=e^{-\tau(x_3+i\Psi(x'))}a_0,\quad V\vert_{\Sigma_0}=0 .
\end{equation}

We multiply any smooth function $a_0$ satisfying (\ref{10}) with a
solution of equation (\ref{lora}) which is supported around the ray
$\{x\vert x_2>0, x_1=0\}$ and is equal to $1$ on this ray. Hence we
can assume that the support of function $a_0$ is concentrated around
this ray and (\ref{zemlanika}) holds true.

Since the Dirichlet-to-Neumann maps of the Schr\"odinger equations
with potentials $q_1,q_2$  are the same there exists a solution to
the following boundary value problem
\begin{equation}\label{zil}
L_{q_2}(x,D)u_2=0\quad \mbox{in}\,\,Q,\quad u_1=u_2\,\,\quad{\mbox
on}\,\,\partial Q, \,\, \mbox{and} \,\, \frac{\partial u_1}{\partial
\nu}=\frac{\partial u_2}{\partial \nu}\,\, \quad{\mbox
on}\,\,\partial Q\setminus \Sigma_0.
\end{equation}
Setting $u=u_1-u_2$ and using (\ref{zil}) we have
\begin{equation}\label{zima}
L_{q_2}(x,D)u=(q_1-q_2)u_1 \quad \mbox{in}\,\,Q,\quad
u\vert_{\partial Q}=\frac{\partial u}{\partial  \nu}\vert_{\partial
Q\setminus \Sigma_0}=0.
\end{equation}

Applying to equation (\ref{zima}) the Carleman estimate
(\ref{solnishko}) we have that there exist constants $C$ and
$\tau_0$ independent of $\tau$ such that

\begin{equation}\label{zebra}
\Vert u e^{-\tau\varphi}\Vert_{H^{1,\tau}(Q)}\le C \quad\forall
\tau\ge \tau_0.
\end{equation}

 Then taking the scalar product in $L^2(Q)$ of  equation (\ref{zil}) with $V$, and using (\ref{zebra}), (\ref{ol}), (\ref{lora1}) we have
  $$ \int_Q(q_1-q_2)u_1Vdx=\int_\Omega uL_{q_2}(x,D)Vdx=O(\frac 1\tau)\quad\mbox{as}\,\,\tau\rightarrow +\infty.
$$
This equality  and (\ref{cor}) imply
$$
\int_Q(q_1-q_2) e^{N(x_3+i\Psi(x'))}aa^2_0 dx=O(\frac 1\tau)
\quad\mbox{as}\quad \tau\rightarrow +\infty.
$$
Therefore
\begin{equation}\label{verka}
\int_Q(q_1-q_2)
e^{N(x_3+i\Psi(x'))}aa^2_0dx=0.
\end{equation}

Setting $p_N(x')=\int_0^L(q_1-q_2) e^{Nx_3}dx_3$  and using
(\ref{zemlanika}) we obtain from (\ref{verka})
\begin{equation}\label{vova}
\int_0^K p_N e^{iNx_2} dx_2=0.
\end{equation}

Indeed, let $r=r_{h}$ be a function such that it is equal $1/2h$ on
the segment $[-h,h]$ and zero otherwise. Denote the solution to
equation (\ref{lora}) given by (\ref{big1}) with the initial
condition $r_{h}$  as $a(h).$  By (\ref{big1}) the function $a_0(h)$
given by formula
$$
a(h)=\left\{ \begin{matrix}\frac{1}{2h} \quad x' \in \Pi_h\\
0 \quad x'\notin \Pi_h,
\end{matrix}\right.
$$ where
$\quad \Pi_h=\{x'\in \Bbb R^2_+\vert x_2\in [0,K]\, -h+\frac{\alpha(
-h)}{\beta(-h)}x_2\le x_1\le h+\frac{\alpha( h)}{\beta(h)}x_2\}.$
Therefore for any fixed $x_2$ from  the segment $[0,K]$ the function
$a(h)$ equals $\frac{1}{2h}$ on the segment $[-h+\frac{\alpha(
-h)}{\beta(-h)}x_2,  h+\frac{\alpha( h)}{\beta(h)}x_2].$ By
(\ref{zombi}) the length of this segment is
$2h+2\alpha'(0)x_2h+o(h).$  We rewrite (\ref{verka}) as
\begin{equation}\label{verka1}
0=\int_Q(q_1-q_2)
e^{N(x_3+i\Psi(x'))}a(h)a^2_0dx=\frac{1}{2h}\int_{\Pi_h}p_N
e^{iN\Psi(x')} a^2_0dx'=\int_0^K\int^{h+\frac{\alpha(
h)}{\beta(h)}x_2}_{-h+\frac{\alpha(
-h)}{\beta(-h)}x_2}\frac{p_Ne^{iN\Psi(x')}a^2_0}{2h}dx_1dx_2.\nonumber
\end{equation}
Applying (\ref{zemlanika}) and using the assumption on regularity of
potentials $q_j$ we have $$ 0=\int_0^K\int^{h+\alpha'(
0)hx_2}_{-h-\alpha'(0)
hx_2}\frac{p_Na^2_0e^{iN\Psi(x')}}{2h}dx_1dx_2+o(1)=
\int_0^K\int^{h+\alpha'( 0)hx_2}_{-h-\alpha'(0)
hx_2}\frac{p_N(0,x_2)e^{iN
x_2}+o(1)}{2h(1+\alpha'(0)x_2)}dx_1dx_2+o(1).$$ This proves
$(\ref{vova}).$

Next we claim
\begin{equation}\label{vovan}
\int_0^K p_N e^{-iNx_2} dx_2=0.
\end{equation}

The proof of (\ref{vovan}) is exactly the same as the proof of
equality (\ref{vova}). The only difference is that instead of the
function $\Psi$ one has to use the function $\tilde\Psi$ defined by
\begin{equation}
\label{vovochka1} \vert\nabla '
\tilde\Psi\vert=1\quad\mbox{in}\,\,\Omega,\quad
\tilde\Psi\vert_{x_2=0}=m,
\end{equation}
\begin{equation}\label{zavtra1}
\tilde\Psi(x_0-t\alpha(x_0)\vec e_1 +t\beta(x_0)\vec e_2)=m(x_0)- t
\quad \forall x_0\in \{x'\vert x_2=0\}\,\,\mbox{and}\,\, \forall
t>0,
\end{equation}
where $\alpha(x_0)= m'(x_0),\beta(x_0)=\root\of{1-\alpha^2(x_0)}.$

From (\ref{vova}) and (\ref{vovan}) setting $N=-i\gamma$ where
$\gamma$ is the real parameter we have
$$
\mathcal R_\gamma (p_{-i\gamma})(\omega, p)=0\quad \forall (\omega,
p)\in S^1\times \Bbb R^1 \quad \mbox{such that}\quad
\{x\vert<\omega,x>=p\}\cap Ch(\Omega\setminus \mathcal O)=\emptyset
.
$$

Applying corollary \ref{zebra1} we have that
$$
p_{-i\gamma}(x')=0\quad \forall x'\in \mathcal O
\quad\mbox{and}\quad \forall \gamma\in \Bbb R^1.
$$
Therefore for any fixed $x'\in \mathcal O$ and any $\gamma$
$$
\int_0^T(q_1-q_2)(x',x_3)e^{i\gamma x_3}dx_3=0.
$$
This equality implies immediately that the function $x_3\rightarrow
(q_1-q_2)(x',x_3)$ on the segment $[0,L]$  is orthogonal to any
polynomials. Therefore $(q_1-q_2)(x)\vert_{\mathcal O\times
[0,L]}=0.$ The proof of the theorem is complete.

 $\blacksquare$


\begin{thebibliography}{99} %










\bibitem{BuU}  A. Bukhgeim and G.\ Uhlmann, \textit{Recovering a
potential from partial Cauchy data}, Comm. Partial Differential
Equations, {\bf 27} (2002), 653--668.

\bibitem{C}  A. P.\ Calder\'on, \textit{On an inverse boundary value
problem,} in \emph{Seminar on Numerical Analysis and its
Applications to Continuum Physics}, 65--73, Soc. Brasil. Mat., R\'io
de Janeiro, 1980.

\bibitem{CIN} M. Cheney, D. Issacson and J.C. Newell,
\textit{Electrical impedance tomography}, SIAM Review, {\bf 41}
(1999), 85-101.


\bibitem{Helg} S. Helgason, \textit{ Integral Geometry and Radon
Transforms,} Springer, Berlin, 2011.

\bibitem{Ik} M. Ikehata, \textit{Inverse conductivity problem in the
infinite slab}, Inverse Problems, {\bf 17} (2001), 437--454.







\bibitem{IUY} O. Imanuvilov, G. Uhlmann and M. Yamamoto,
\textit{The Calder\'on problem with partial data in two dimensions},
J. Amer. Math. Soc., {\bf 23} (2010), 655-691.

\bibitem{IUY1} O. Imanuvilov, G. Uhlmann and M. Yamamoto,
\textit{Partial Cauchy data for general second order elliptic
operators in two dimensions}, arXiv:1010.5791v1 to appear in
Publications of the  Research Institute for Mathematical Sciences.


\bibitem{IY} O. Imanuvilov and M. Yamamoto,
\textit{Inverse boundary value problem for Schr\"odinger equation in
two dimensions}, SIAM J. Math. Anal., {\bf 44,} (2012), 1333-1339.

\bibitem{I} V. Isakov,
\textit{On uniqueness in the inverse conductivity problem with local
data}, Inverse Problems and Imaging, {\bf 1} (2007), 95-105.

\bibitem{KSU} C.~Kenig, J.~Sj\"ostrand and G.~Uhlmann,
\textit{The Calder\'on problem with partial data}, Ann. of Math.,
\textbf{165} (2007), 567--591.

\bibitem{KLU} K. Krupchyk, M. Lassas and G. Uhlmann
{\it Inverse problems with partial data for a magnetic Schr\"odinger
operator in an infinite slab and on a bounded domain,} Comm. math.
Phys., {\bf 312}, (2012),  87--126.

\bibitem{Kur} A. Kurusa {\it Translation-invariant Radon transforms,}
Math. Balcanica, {\bf 5,} (1991), 40-46.

\bibitem{RG} R.G. Novikov, \textit{ A mutidimensional inverse
spectral problem for the equation $-\Delta \psi+(v(x)-E(v(x))\psi$}
Functsional Anal i Prilozhen, {\bf 22} (1988), 11--22.


\bibitem{Nov} R.G. Novikov, \textit{New global stability estimates for
the Gel'fand-Calderon inverse problem,} Inverse problem \textbf {27}
(2011),  015001.

\bibitem{LG} X. Li, G. Uhlmann
\textit{Inverse problems with partial data in slab,} Inverse Prob.
Imaging, {\bf 4,} (2010), 449--462.





\bibitem{N} A.~Nachman, \textit{Global uniqueness
for a two-dimensional inverse boundary value problem}, Ann. of
Math., \textbf{143} (1996), 71--96.








\bibitem{SU} J.~Sylvester and G.~Uhlmann,
\textit{A global uniqueness theorem for an inverse boundary value
problem}, Ann. of Math., \textbf{125}  (1987), 153--169.

%






\end{thebibliography}
\end{document}